# Effects of Deception in Social Networks

Gerardo Iñiguez [1], Tzipe Govezensky [2], Robin Dunbar [3],
Kimmo Kaski [1] and Rafael A. Barrio [4, 1]

[1] Department of Biomedical Engineering and Computational Science, Aalto University School of Science, FI-00076 AALTO, Finland
[2] Instituto de Investigaciones Biomédicas, Universidad Nacional Autónoma de México, 04510 México D.F., Mexico
[3] Department of Experimental Psychology, University of Oxford, OX1 3UD, United Kingdom
[4] Instituto de Física, Universidad Nacional Autónoma de México, 01000 México D.F., Mexico

**Honesty plays a crucial role in any situation where organisms exchange information or resources. Dishonesty can thus be expected to have damaging effects on social coherence if agents cannot trust the information or goods they receive. However, a distinction is often drawn between prosocial lies ("white" lies) and antisocial lying (i.e. deception for personal gain), with the former being considered much less destructive than the latter. We use an agent-based model to show that antisocial lying causes social networks to become increasingly fragmented. Antisocial dishonesty thus places strong constraints on the size and cohesion of social communities, providing a major hurdle that organisms have to overcome (e.g. by evolving counter-deception strategies) in order to evolve large, socially cohesive communities. In contrast, "white" lies can prove to be beneficial in smoothing the flow of interactions and facilitating a larger, more integrated network. Our results demonstrate that these group-level effects can arise as emergent properties of interactions at the dyadic level. The balance between prosocial and antisocial lies may set constraints on the structure of social networks, and hence the shape of society as a whole.**

1. Introduction

Reciprocity plays an important role in the emergence of cooperation both in real life and in theoretical models [1–7]. Though rarely explicitly operationalized in models of cooperation, trust plays a central role in this process because it underpins individuals' willingness to engage in exchange interactions [6,8–10]. In this context, deception becomes important because it threatens to undermine the delicate balance on which trust-based exchange relationships are based. In the evolutionary literature, there has been a longstanding concern about the fact that, since deception is potentially advantageous to signallers competing for resources, the ease with which it can evolve is likely to destroy the stability of trust-based relationships [11,12]. Much of the extensive behavioural and evolutionary literature on this topic has been concerned with identifying mechanisms for detecting and policing deception and hence the evolutionary stability of honest communication, and thus relates closely to the broader literature on the management of free-riders [13]. Most modelling approaches have used evolutionary game theory or evolutionarily stable strategy (ESS) dynamics [13–16], and many have shown that some balance between honesty and deception yields an evolutionarily stable outcome, with the ratios between the two states generally dependent on the precise form of the payoff matrix [16]. Much of the literature in animal behavioural ecology (reviewed in [11,12]) provides empirical support for these theoretically derived conclusions.

This approach has been predicated on the (reasonable) assumption that lying is destructive because it destabilises the relationship of trust between pairs of interacting individuals. As a result, most models have considered panmictic populations in which individuals interact indiscriminately with each other. The same has largely been true of the extensive animal behavioural ecology literature (reviewed in [11,12]), as well as the much more limited empirical literature on this topic in economics [17,18], psychology [19,20] and even robotics [21]. However, the populations of most highly social species (many birds and mammals, but especially humans and other primates) are highly structured [22–24] and the implications of trust (or its converse, deception) for the structure of social networks in these cases has not been explored in any detail [25–27].

Such issues may be especially intrusive in intensely social taxa like anthropoid primates (including humans), equids, elephants and the dolphin family [28], In these taxa, an individual's fitness is the product of two separate components, namely the consequences of its own decisions and the effects that its neighbours' decisions have on its fitness through group-level effects (as reflected in Hamilton's classic concept of *neighbourhood modulated fitness* [29,30]). In these cases, individuals gain fitness by sharing in the collective behaviour of the community to gain protection from predators, defend resources or exchange information on the location of ephemeral resources. Group-level effects of this kind often depend not only on the size of the group but also on how well the group is integrated (or bonded) [31]. Bondedness reflects the level of cohesion within the network, its degree of connectedness and its integrity over time. In these cases, network integrity becomes a critical component of the individual's inclusiveness fitness, and individuals may lose fitness when actions such as lying adversely affect network integrity because they reduce trust.

Some sense of the scale of the issue is provided by the fact that lying appears to be surprisingly common among humans: a randomized representative sample of 1,000 Americans recorded an average of 1.65 lies per day (equivalent to ~550 per person per year), although the distribution was highly skewed (23% of all lies were told by 1% of individuals, and half of all lies were told by 5% of the respondents) [32]. A number of other studies have reported frequencies between 0.6-2.0 lies per day, with lies being less common in face-to-face interactions than in telephone interactions [33–35]. Many of these may be trivial or prosocial lies (lies to avoid causing offence or lies to protect a third party [36]). However, in one study, 92% of respondents claimed to have lied at some time to a partner [37], while another study found that 60% of women and 34% of men had lied to obtain sex [38], suggesting that even modest lies may often be used for purely personal gain. Equally surprisingly, subjects in experimental games were two to three times more likely to cheat if they gained a benefit by doing so [39,40].

Given this, it is perhaps not surprising that humans are sensitive to both the magnitude and the intention of lies: large actual or potential gains for the liar are more likely to be frowned on than minimal gains [17,36], even in one-shot games with no possibility of future interaction [41]. One possible explanation is the risk of an unfavourable social reputation and/or the implementation of strict social sanctions against lying and deception [42,43], but a number of studies report that the maintenance of social cohesion or relationship stability was viewed as an acceptable motive for lying and, more broadly, that prosocial lies are rated as more acceptable than antisocial lies [36,44].

In this paper, we focus on the impact that lying has on the coherence and structure of social networks and use a recent opinion dynamics model [45] to explore this. Conventional approaches to collective action typically focus on the direct payoffs of the decisions that individuals make. In contrast, we focus on the extent to which such decisions affect network cohesion and the indirect effects that this has on individual fitness. This approach shares with the models developed by Dávid-Barrett & Dunbar [23,46,47] a focus on achieving behavioural synchrony as an intermediate goal necessary for maximising fitness, in line with Hamilton's conception of neighbourhood modulated fitness [29,30]. More importantly, it bypasses the conventional collective action dilemma created by free-riders because it is not possible to free-ride: one is either in the group or one is not, and one benefits accordingly [46].

We are concerned here not with explaining how deception is counteracted by detection mechanisms so that animals can be social (the principal focus of much of the evolutionary and economics literature), but rather with exploring whether, and to what extent, deception is destructive of sociality. By sociality, we do not mean the dyadic relationships that have been the focus of most empirical and modelling studies hitherto, but the capacity to build complex, integrated social networks of the kind found in socially more complex species such as primates, elephants, and delphinids among the mammals [22], and of course humans [24].

Social psychologists commonly characterize the degree of human deception by distinguishing four types of lying: (i) *Prosocial*, lying to protect someone, or to benefit or help others; (ii) *Self-enhancement*, lying to save face, avoid embarrassment, disapproval or punishment or gain an advantage (these lies are not intended to hurt anyone, rather they benefit the self); (iii) *Selfish*, lying to protect oneself at the expense of another and/or to conceal a misdeed; and (iv) *Antisocial*, lying to hurt someone else intentionally [17]. In this study, we will distinguish between two general types of lies: prosocial lies of type (i) and antisocial lies of types (ii)-(iv). Of these two general types, it is widely recognised that antisocial lies are destructive of relationships because they are selfish (the liar gains fitness at the expense of the target of the lie), whereas prosocial lies ('fibs') help to keep relations in good condition [26,48].

Our analysis will focus explicitly on dyadic patterns of lies (lies to a specific individual), because we wish to explore the extent to which lies at the dyadic level influence the structure and cohesion of entire networks. For this reason, we hold back from introducing an evolutionary dimension to the model so that we can assess, as a first step, the impact that different liar phenotypes have on network structure. Lies can also be conceived of at the community level (a lie is broadcast to the entire group simultaneously, as studied in many cultural diffusion models [49]), though it is possible that this requires language. An example of a community-level lie might be the promulgation of a particular religious belief (e.g. a god that punishes misdemeanours [50]) which, while harmless enough in its own way, may be untrue, yet belief in the lie may result in greater community cohesion to the wider benefit of the members. Lies at this scale impact on social cohesion in a more direct and obvious way, and are thus less interesting for present purposes than whether or not lies at the dyadic level have any effect on the cohesion of extended networks. The latter is important because while a white lie may be beneficial for a dyadic relationship, it could be so at the expense of other relationships or other group members, and hence at the same time disadvantageous for the group as a whole.

## 2. Methods

To study the social implications of deception we use an agent-based model of opinion formation in coevolving social networks [49,51], in which a continuous state variable $x_i(t)$ representing the instantaneous opinion of agent $i$ at time $t$ about an issue or topic is bounded between $-1$ and $+1$ (i.e. total disagreement and total agreement, respectively). Here we represent society by a network of $N$ agents, each connected to $k_i$ other agents and thus giving rise to a structured society, rather than a panmictic population. In a previous model [45], we proposed the basic dynamical equation for the opinion state variable of an individual to be as follows,

$$\frac{\partial x_i}{\partial t} = f_s(i)|x_i| + \alpha_i f_l(i), \qquad (1)$$

where the first term on the right-hand side represents a "short-range" exchange of information or opinion with neighbouring agents, and the second term stands for the "long-range" influence of the overall opinion of the network on agent $i$, which in turn is modulated by its attitude parameter $\alpha_i$ (for or against the overall opinion $f_l(i)$ of the crowd, depending on whether the attitude is positive or negative). This system evolves with a characteristic time scale $dt$, referred to as the "transaction time". With this approach we showed that well-connected opinion groups emerge naturally when opinion change rises during homophilic interactions, individual attitudes affect society's overall mood, and opinion homogeneity influences network evolution [45].

While we consider this dynamical state equation approach generally valid and useful, we need to take into account that in real social networks links often have different weights and the exchange of information between two individuals is not necessarily truthful. In order to include these two key aspects of agents in a social network into the previous model, we introduce a second state variable $y_i$ to represent the perception of the true opinion $x_i$ of agent $i$ by other agents. This means that the true opinion is private, while the public opinion can be written as,

$$y_i = \frac{1}{k_i} \sum_{j \in m(\ell=1)}^{k_i} w_{ji}, \qquad (2)$$

where $w_{ji}$ is a matrix element proportional to the amount of information flowing from agent $i$ to neighbouring agent $j$ at the time of interaction $t$, and $k_i$ is the degree of agent $i$, i.e. the size of the set $m(\ell = 1)$ comprising its direct neighbours in the network. Then the first term of Eq. (1) can be written as,

$$f_s(i) = \sum_{j \in m(\ell=1)}^{k_i} w_{ij}, \qquad (3)$$

Moreover, the overall opinion now depends on the public variables of agents further away in the network,

$$f_l(i) = \sum_{j \in m(\ell=2)}^{m(\ell_{max})} \frac{1}{\ell} y_j, \qquad (4)$$

where $\ell = 2, \ldots, \ell_{max}$ is the distance between agent $i$ and non-neighbouring agent $j$. In this model an honest interaction between two agents implies $w_{ij} = x_j$, while an act of deception is construed as manipulating the information $w_{ij}$ from true to false, as detailed below.

In order to follow the dynamical changes on the weights of the links, we introduce an equation for the temporal evolution of the adjacency matrix $A_{ij}(t)$,

$$\frac{\partial A_{ij}}{\partial t} = DT(x_i(t), y_i(t), x_j(t), y_j(t)), \qquad (5)$$

where the slope $T$ is a function of the four opinion variables involved in each link, and the parameter $D$ sets the time scale for link growth. Here we assume that links get stronger if a) their values are close to the extremes $\pm 1$ (agents have firm convictions), and b) the difference between $x$ and $y$ is small (honesty strengthens relations). A suitable linear combination that agrees with these assumptions is,

$$T = \left( \left| \frac{(x_i + x_j)}{2} + \frac{3(y_i + y_j)}{2} \right| - 1 \right). \qquad (6)$$

Near the point of total conviction (i.e. $x, y \sim \pm 1$), this expression can have four different values:
1) $T = 3$, if there is a truthful convergence of opinion ($x_i = x_j = y_i = y_j = \pm 1$).
2) $T = 2$, if there is a truthful divergence of opinion ($x_i = 1 = -x_j$ and $y_i = y_j = \pm 1$).
3) $T = 0$, if there is a truthful convergence of opinion but public disagreement ($x_i = 1 = x_j$ and $y_i = -y_j$).
4) $T = -1$, if there is a divergence of opinion involving deception ($x_i = 1 = -x_j$ and $y_i = -y_j$).

If the slope $T$ is negative, the link weakens and eventually its weight becomes zero, effectively breaking the connection between agents.

Given that in the model the probability that links between agents will be broken is proportional to the difference in their opinions (as encoded in the expression for $T$), a lie implying $w_{ij} \sim y_i$ is beneficial for agent $j$ in that the apparent similarity prevents the link with agent $i$ from being broken. We consider this act of deception to be *prosocial*, since agent $i$ also benefits from having such a neighbour (in the sense that the link is likely to be reinforced). On the other hand, a value $w_{ij} \sim -y_i$ is bound to create tension between differing opinions and thus give rise to *antisocial* deception (since the opinion difference with the neighbour will cause the link to weaken). Overall, we choose to write the information flow $w_{ij}$ as,

$$w_{ij} = \begin{cases} x_j, & \text{if } A_{ij} \neq 0 \text{ and } A_{ij}^2 \neq 0 \\ \tau x_j \pm (1-\tau) y_i, & \text{otherwise} \end{cases}, \qquad (7)$$

where the honesty parameter $\tau$ is an average measure of how much individuals, given the cultural traits of their society, reject lying and the sign of the term including $y_i$ distinguishes between pro- and antisocial lies. In other words, agents always tell the truth to neighbours in a triad (the minimal component of a social subgroup), and tell a partial truth to the rest. Here $\tau$ is taken to vary from 0 (dishonesty) to 1 (honesty). In Figure 1 the three possible choices of pairwise interactions, namely to be honest or to lie pro- or antisocially, are illustrated.

This development of social structure is based on two main assumptions: (i) honest relations between agents with similar opinions growing in strength as time goes by, while deceptive relations with divergence of opinion weaken and eventually break the link between them; and (ii) agents with fewer links are eager to make new links to avoid becoming marginalized.

One of the key features of models like the present one is that the pace at which new social network links are created is much slower than the typical time scale for the state variable dynamics. This reflects the fact that people do not change their relationships very fast. Therefore, we set a time scale $g$ for a network rewiring process that conserves the total number of links. To do this we assume that, after $gdt$ time steps of the transaction dynamics, agents seek for new relationships. At these times, we count the number $n_c$ of weighted links that have been set to zero [due to the dynamics of Eq. (5)] and add the same number of links. This is realised by first examining the resulting network and assuming that agents with fewer links are more eager to make new ones, in particular the agents that are left without links and have become marginalised. Then we order agents by ascending number of links (i.e. by their degree) and choose agents to be rewired in the order of such list. The number of cut links $n_c$ is related to the number $n$ of agents chosen to be considered for a new link by $n_c = n(n-1)/2$, where the right-hand side is the number of links in a fully connected subgraph of $n$ agents. This relation can be rewritten as follows,

$$n = \left(1 + \sqrt{1 + 8n_c}\right)/2. \tag{8}$$

The $n$ chosen agents create all previously non-existing links among themselves; if the number of links created is still below $n_c$, we connect random pairs of agents until reaching $n_c$. In this way the rewiring process keeps the total number of links constant, while promoting reconnection for marginalised agents.

It should be noted that the present model contains three different time scales: 1) *dt* is the typical time interval between "transactions" (that is, the dynamics of the opinion state variable), 2) *Ddt* is the time scale with which a link changes its weight, and 3) *gdt* represents the typical time required to make new acquaintances.

We examine various deceptive scenarios by solving the equations of our model numerically, using a simple Euler method with fixed time step. We perform extensive simulation runs starting from a random Erdős-Rényi network [52] of size $N$ with average degree $\langle k_0 \rangle$ and a uniform distribution of initial random opinions in the interval $(-1, +1)$. The evolution of the opinions of agents and the coevolution of their network is followed for either prosocial or antisocial deception and for a constant honesty parameter value $\tau$. To allow for a stationary final state in which the properties of the network do not change anymore, honest opinions with magnitude greater than one become fixed, i.e. $x_j = \pm 1$ and are considered as the final decision state of the population.

### 3. Results

In Figure 2 we illustrate the effects of pro- and antisocial lying in our modelled social network for either semi-honest ($\tau = 1/2$) or totally dishonest ($\tau = 0$) societies, as compared to the structure of a completely honest network (corresponding to $\tau = 1$ and shown at the top of the figure). Here we see that, for prosocial deception in a semi-honest society, the network becomes separated into two large communities of roughly the same size, with strong links within social groups and weak links between them. We note that the connections between the two large clusters are due to undecided agents with $|x_j| < 1$. In the case of a totally dishonest society with prosocial agents, the two large clusters become fragmented into small sub-communities that are basically cliques of individually honest agents, with individually dishonest agents providing weak links between groups. Here we consider agent $j$ to be individually honest (dishonest) if its individual deception $d_j = |x_j - y_j|$ is smaller (larger) than the average openness in the social network, $d = \langle d_j \rangle$. The effect of fibs is in this way clear, as it facilitates the formation of small, tightly connected communities. When the deception process dominates, groups become smaller and more tightly linked, and the numerous weak links between agents that do not belong to small communities turn out to be exclusively between dishonest agents.

Next we examine the effects of antisocial lies, that is, the case in which a deceptive information flow takes the form $w_{ij} = \tau x_j - (1 - \tau)y_i$. In Figure 2, the two panels in the bottom row show the effect of this form of information flow on the topological structure of the final network. Note that the social network corresponding to an honesty parameter $\tau = 1/2$ is much less connected than its counterpart in the case of prosocial lies, and shows small communities of honest agents sharing the same opinion but being weakly connected by dishonest agents. This effect is even more pronounced when the amount of lying increases ($\tau = 0$). In this extreme case of universal antisocial lying, almost all agents become completely isolated and the network structure is destroyed. These results suggest that antisocial lies may destroy the cohesion of a social network, whereas prosocial lies may, paradoxically, enhance its cohesiveness.

In the right hand corner of Figure 2 we have investigated the dependence of the number of undecided agents on the honesty parameter $\tau$ for prosocial and antisocial behaviour, by monitoring the number of agents $j$ that have not reached a definite decision (i.e. with $|x_j| < 1$). Here we see that for the prosocial case the number of undecided agents decreases slightly for an increasing volume of lying, suggesting that prosocial lies actually help people make up their minds. In contrast, antisocial deception increases the frequency of undecided agents in the network, which shows that misleading information may make the process of decision-making very difficult. In the bottom right hand corner of Figure 2 we investigate the dependence of the local clustering coefficient, averaged over all agents, on the honesty parameter $\tau$ for prosocial and antisocial behaviour. The local clustering coefficient is the probability that two randomly selected neighbours of an agent are connected [53], and its average over the whole network serves as an index of interconnectedness for the population of agents. We observe that the clustering coefficient is quite insensitive to prosocial lies, while it increases enormously with the amount of antisocial lies. This result suggests that in a society where one is honest only with close friends and family, and totally antisocial with the rest, the structure of society becomes fragmented into small but well-connected groups opposing each other.

In Figure 3, we investigate the effect of the individual deception $d_j$ of agents on two local network characteristics in a dishonest ($\tau = 0$) society, both for pro- and antisocial lies. One of these is the weighted clustering coefficient, an extended version of the local clustering coefficient where the weight $A_{ij}$ is taken into account [54]; the other is the betweenness centrality, a measure of the number of shortest-paths going through an agent [55] reflecting its importance in the social network. Here we see that honest agents have low betweenness centrality but high clustering, forming small and tight communities with truthful interactions. Moderately deceptive agents maximize their betweenness and become very central in the network, acting as bridges between groups, since their clustering is seen to decrease rapidly. At the other extreme, totally dishonest agents become marginalized, with no centrality or triads around them. This qualitative behaviour holds for both kinds of deception, although prosocial lying typically produces larger betweenness and lower clustering than antisocial lying.

## 4. Discussion

Although it is widely recognised that deception can have an adverse effect on social cohesion, and might hence lead to the fragmentation of social networks, our results suggest that in communities dominated by honest individuals, dishonest agents can sometimes act as weak links and serve as communication paths between clusters of honest agents. More importantly, however, there is a difference between prosocial and antisocial deception, since prosocial lies give rise to a rich network structure with tight communities connected by weak links (either by liars or undecided agents), while antisocial lies tend to radically disconnect the network. Recall that our definition of prosocial lies ('white lies' or fibs) is that they benefit the recipient at a modest cost to the actor, thereby preserving the dyadic relationship between them – in a context where an antisocial lie would threaten the stability of the relationship. Lies of this kind seem to occur commonly in human interactions: there is experimental evidence, for example, to suggest that people may lie or cheat to ensure greater equity of outcomes in economic games [56]. This may be important for maintaining relationship equilibrium in networks since a highly inequitable distribution of payoffs can be very disruptive (especially in traditional small scale societies [57]). This suggests that there may be an important distinction to be drawn between prosocial and antisocial lies that has so far escaped the attention of evolutionary biologists modelling the evolution of deception.

Prosocial lies at the community scale are likely to depend on language, and will thus be unique to humans. Examples might include mass indoctrination via religious or political proselytising: these commonly percolate through social networks by cultural transmission, but their function is explicitly to facilitate within-community cohesion [58]. However, below this is a separate level at which lying might serve a function, namely the level of dyadic interactions, which themselves create the networks that give a community its particular shape. Many studies of ethnographic societies report that, in small scale societies, people are prone to answering questions in the way they think the questioner would like to hear rather than by telling the truth (unexpected answers are considered rude and hence potentially socially divisive). By ensuring that everyone in a community abides by the same rules of behaviour, it is not difficult to see how cases like this might ensure that groups continue to provide the kinds of fitness benefits for individuals that we noted earlier. Our concern here has been with whether prosocial lies at the *dyadic* level might also feed back into community level effects by influencing network structure and cohesion in a context where an individual is influenced by how effectively the community works together to solve core ecological problems.

In this respect, our apparently simple model, with basic social interactions represented by minimal rules, allows us to draw some conclusions about the role of deception in social interactions. In particular, our findings are reminiscent of the observation that humans commonly draw a distinction between prosocial and antisocial lies, and are more willing to accept prosocial or 'polite' lies in appropriate social circumstances [20,59,60]. One may wonder why, if lies are universally considered as bad, there is no human society in which they are totally absent. Although we are taught from an early age that lying is unacceptable, children nonetheless learn to lie wisely instead of eschewing it completely. The results of our study suggest that not all lies are bad or necessarily socially destructive; in fact, it seems that some lies may even enhance the cohesion of the society as a whole and help to create links with other people. In effect, some kinds of lies might actually be essential to the smooth running of society, and if so, the balance between pro- and antisocial lies may be crucial in shaping social structure. Curiously, our results seem to suggest that totally honest interactions prevent diversity in the form of highly clustered social subgroups, while totally dishonest interactions result in the destruction of the social network. An intermediate level of deception may thus be optimal to perform certain social functions (such as protection from outsiders, resource defence and information exchange), especially when social systems are complex and multi-layered.

While the distinction between prosocial and anti-social lies is obvious in humans, it is not at all clear whether prosocial lies in this sense exist among nonhuman animals. This might be because the ability to distinguish between prosocial and antisocial lies and intentionally engage in the first in order to maintain network cohesion implicitly depends on advanced mentalising (theory of mind) competences [61] as well as language. If so, then our model may be explicitly relevant only to humans, since we are the only species capable of true mentalising [62]. Our model bears some similarities to that of McNamara et al. [63,64], which essentially explores the evolutionary stability of different personality types in the context of trust-based relationships – of which the propensity to lie or tell the truth might be a relevant exemplar. Their model focuses on the way individuals monitor their interactions with others, whereas our model in some sense takes this for granted and asks whether beneath this there may still be room for a less destructive form of 'untrustworthiness', one that might help maintain the integrity of dyadic networks in contexts that might otherwise destabilise the relationship.

Nonetheless, it is worth asking what prosocial lies might look like in nonhuman species. Possible candidates include distraction displays such as the feigned brooding and broken wing displays observed in many shorebirds (plovers, killdeer and lapwing) [65–67] and fish (e.g. stickleback: [68,69]), although in all these cases deception seems to relate mainly to the deceiver's immediate fitness interests (i.e. offspring survival) rather than the wider interests that we suggest prosocial lies serve in facilitating community cohesion. More plausible examples might be animals that give false predator alarms when another individual moves too far away from their group (possible examples have been noted in vervet monkeys [70]). Other cases of deception in primates that have been interpreted as tactical deception [71] (including Kummer's iconic description of a female *Papio hamadryas* baboon using a rock to prevent her harem male seeing that she was grooming with a male from a neighbouring harem, or female chimpanzees suppressing their loud post-copulation calls when they know a dominant male [72] or female [73] is within earshot) could be interpreted as prosocial rather than pure self-interest if the aggression that would normally follow the discovery of such behaviour had destabilising effects on the group as a whole. In contrast to the case in humans, such examples would not be intentional prosocial deception, but functionally they may have the same consequences. If so, they might allow mild prosocial deception (the recipient benefits) to be distinguished from serious deception (the recipient loses out), and so provide a foundation from which truly prosocial lies might evolve in species that have the cognitive ability to act in this way.

Our analyses suggest that prosocial forms of deception may only arise in the context of socially complex species that have multilevel social systems based on structured networks [22], where higher-level groupings provide additional group-level fitness benefits [47,74] that make ensuring the continued cohesion of these grouping levels functionally important. Nonetheless, given that these two forms of deception can be distinguished in at least one species (i.e. humans), it raises the question of whether prosocial deception is an evolutionary precursor to selfish, antisocial deception, or an emergent property of the latter that evolved as a benefit once deception was well entrenched and the species had sufficient cognitive capacity to calculate the consequences of behaviour.

Our aim in this paper has principally been to address the consequences for network structure of having populations made up of different behavioural phenotypes (honest agents vs different kinds of liars). We have deliberately avoided, for these purposes, any attempt to explore the evolutionary processes whereby populations achieve an evolutionary stable distribution of phenotypes. Instead, we have taken the distribution of phenotypes to be fixed and asked how this affects network integrity and connectedness. In effect, we view the traits as exogenous. An important follow-up study, and one we are actively exploring, is to investigate the way in which a society evolves a distribution of phenotypes (i.e. the traits are considered as an endogenous component of the system and change in frequency as a result of selection). Establishing, as we have done here, that a prosocial liar phenotype might be beneficial in terms of network connectedness has been an important necessary first step on the way to this objective since, all else being equal, we would not have expected it to be the case, and hence might never have considered it as an option.

**Acknowledgements:** G.I. and K.K. acknowledge support from EU's FP7 FET Open STREP Project ICTeCollective No. 238597. R.D.'s research is funded by a European Research Council Advanced grant. R.A.B. acknowledges support from Conacyt project No. 179616.

**Figures:**

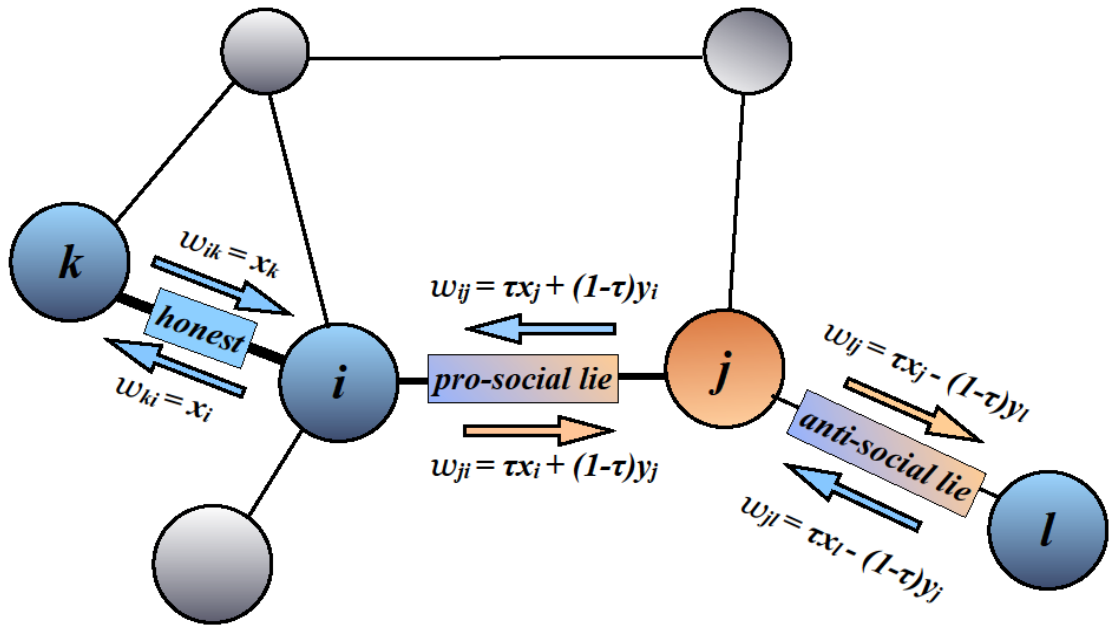

**Figure 1.** Model of opinion dynamics in a social network consisting of honest interactions or exchanges of either prosocial or antisocial lies between individuals. In an honest exchange (inside a triad, for example), the flow of information between agents $k$ and $i$ matches their true opinions. In the case of prosocial lying, agents $i$ and $j$ try to mimic each other's public opinions. In the third case, agents $j$ and $l$ lie antisocially to oppose each other's perceived view on the subject.
17

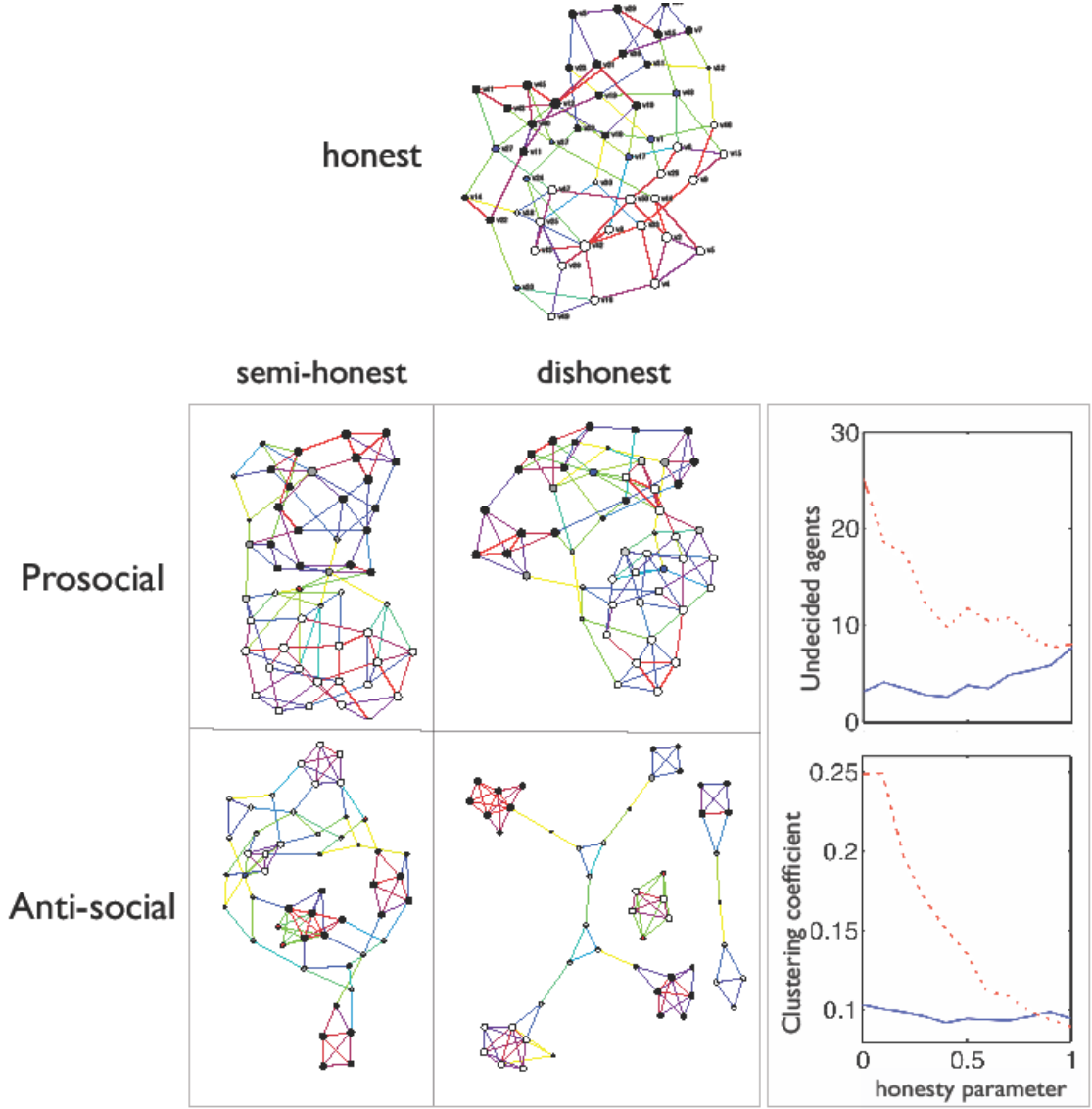

**Figure 2.** Comparison of honest, pro- and antisocial behaviour. On the top and left: Final network structure (starting from the same random network with $N = 50$ and $\langle k_0 \rangle = 6$) for an honest ($\tau = 1$) dynamics, as well as for semi-honest ($\tau = 1/2$) and completely dishonest ($\tau = 0$) dynamics where agents are allowed to lie pro- or antisocially. Node colours are white/black ($x_j = \pm 1$) and light/dark grey ($x_j = \pm 1$, $d_j < d$). The size of nodes is proportional to their degree and labels in the honest network are arbitrary. The width of the links is proportional to their weight. Observe that most links between communities involve undecided agents. On the right: Effect of a varying honesty parameter on the final number of undecided agents and the clustering coefficient of the network for prosocial (continuous line) and antisocial (dashed line) lies. A decrement in $\tau$ corresponds to an increment of the number of lies. Curves are averaged over 10 realisations with $N = 200$. All results correspond to $dt = 0.002$, $D = 3$ and $g = 100$.

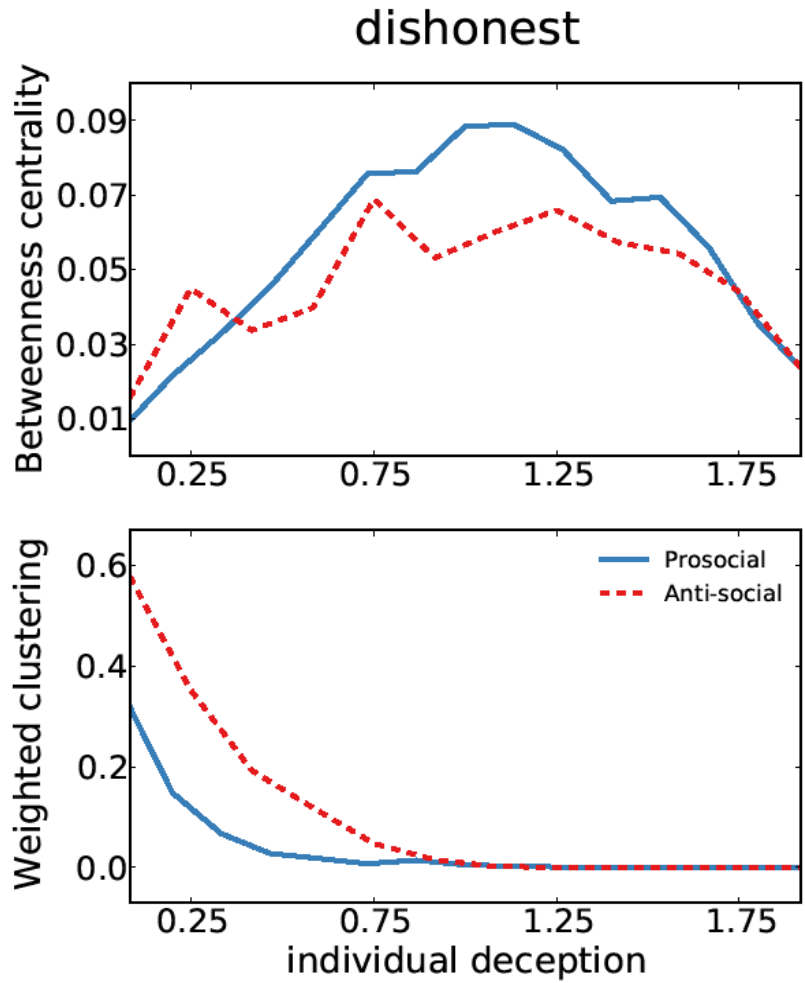

**Figure 3.** Average effect of individual deception on an agent's betweenness centrality (top) and weighted clustering coefficient (bottom), for prosocial (continuous line) and antisocial lies (dashed line) and in the case of a dishonest ($\tau = 0$) society. Typically honesty is related to high clustering, as intermediate deception is to high betweenness centrality values. All curves are averaged over $10^3$ realisations with $N = 200$, $\langle k_0 \rangle = 5$, $dt = 0.002$, $D = 1$ and $g = 100$.